\documentclass[useAMS, usenatbib, usegraphicx, onecolumn]{mn2e}
\usepackage{times}
\usepackage[fleqn, tbtags]{amsmath}
\usepackage{ifthen}
\newboolean{grey-scale}
\setboolean{grey-scale}{false}
\renewcommand{\vec}{\bmath}
\newcommand{\ud}{\rmn{d}}
\newcommand{\satellite}[1]{{\it{#1}}}
\begin{document}
\title{Baryon Oscillations and Dark-Energy Constraints from Imaging Surveys}
\author[Derek Dolney et al.]{
  Derek Dolney$^1$\thanks{E-mail: dolney@astro.upenn.edu;
                                  bjain@physics.upenn.edu;
                                  takada@astr.tohoku.ac.jp},
  Bhuvnesh Jain$^1$\footnotemark[1],
  Masahiro Takada$^2$\footnotemark[1]\\
  ${}^1$Department of Physics and Astronomy, University of Pennsylvania,
  209 S 33rd St, Philadelphia, PA 19104-6396, USA\\
  ${}^2$Astronomical Institute, Tohoku University, Sendai, 980-8578, Japan}
%
%
\pagerange{\pageref{firstpage}--\pageref{lastpage}} \pubyear{2004}
\maketitle
\label{firstpage}
\begin{abstract}
  Baryonic oscillations in the galaxy power spectrum have been studied
  as a way of probing dark-energy models. While most studies have
  focused on spectroscopic surveys at high redshift,
  large multi-color imaging surveys have already been planned for
  the near future. In view of this, we study the prospects for
  measuring baryonic oscillations from angular statistics of galaxies
  binned using photometric redshifts. We use the galaxy bispectrum in 
  addition to the power spectrum; this allows us to measure and 
  marginalize over possibly complex galaxy bias mechanisms to get 
  robust cosmological constraints. In our parameter estimation
  we allow for a weakly nonlinear biasing scheme that may evolve with 
  redshift by two bias parameters in each of ten redshift bins. We find that
  a multi-color imaging survey that probes redshifts beyond one can give
  interesting constraints on dark-energy parameters. 
  In addition, the shape of the primordial
  power spectrum can be measured to better accuracy than with the CMB
  alone. We explore the impact of survey depth, 
  area, and calibration errors in the photometric redshifts on
  dark-energy constraints.
\end{abstract}
\section{Introduction}
Recent cosmological observations have provided 
strong evidence that a dark-energy
component, such as the cosmological constant, comprises as much as
70 per cent of the total energy density of the universe
(\citealt{Perlmutter1999, Riess1998, Riess2001, Tonry2003, SpergelEtAl2003}).
Characterizing the nature of the dark energy and its possible
evolution has become a central goal of empirical work in cosmology. 

Galaxy surveys measure the clustering statistics of galaxies
as a function of scale and redshift. The galaxy power spectrum can
be compared to the CMB power spectrum to constrain the growth of
structure. However the amplitude of the galaxy power spectrum 
depends on the biasing of the particular galaxy sample; one therefore
needs to exercise care in using the full power spectrum for 
cosmological inferences (e.g. \citealt{Tegmark2004, Pope2004}). 
The shape of the power spectrum has been regarded as more robust
to biasing effects. 

The baryon oscillations in the galaxy power spectrum are imprints from
acoustic oscillations in the early universe, prior to recombination. 
The same physics produces the dramatic peaks and troughs seen in
the CMB power spectrum. Their physical scale is set by the 
sound horizon at recombination, which has been 
determined accurately from CMB data (\citealt{SpergelEtAl2003}). The
baryon oscillations in the matter and galaxy power spectra are much
weaker features
because the dark matter which dominates the mass density did not
participate in the acoustic oscillations. The oscillations are at the
level of a few tens of a percent variation about the smoothed power
spectrum. With a survey of sufficient size, these features can
be measured accurately. Since the physical length scale of the baryon 
oscillations is known from the CMB measurement of the sound horizon, 
a measurement of their apparent size in redshift or angle space
leads to a measurement of purely geometric quantities: the Hubble 
parameter and the angular diameter distance, respectively. 
We will be concerned with the relation between physical size
and angular diameter distance: $r_\rmn{physical} = d_\rmn{A}(z)
\Delta\theta$, where $d_\rmn{A}$ is the angular diameter distance and 
$\Delta\theta$ is the measured angular scale. This relation can
be used for a statistical quantity as well; for the power spectrum 
it means that a measurement of an angular wavenumber $l$ and its
relation to the physical wavenumber $k$ yields $d_\rmn{A}(z)$. We 
describe in the next section how $d_\rmn{A}$ constrains models of dark
energy. 

To measure baryon oscillations, many authors have considered galaxy 
surveys over different redshift ranges (\citealt{Eisenstein2003,
 BlakeGlazebrook2003, Linder2003, HuHaiman2003, SeoEisenstein2003,
 MatsubaraSzalay2003}). For spectroscopic
redshift surveys, the tangential and radial components are considered
separately since the latter is subject to redshift distortions. 
Current redshift surveys can map large enough volumes at
redshifts well below 0.5. It is a great advantage to probe higher
redshifts since additional baryon oscillations can then be measured within 
the linear regime of clustering (the linear regime extends to smaller physical
scales at high redshifts). With future redshift surveys, such as 
the proposed KAOS\footnote{http://www.noao.edu/kaos/} survey, such a
measurement would be possible.

Multi-color imaging surveys are already in progress, 
e.g. the SDSS\footnote{http://www.sdss.org/}, the CFH
Legacy\footnote{http://www.cfht.hawaii.edu/Science/CFHLS/} survey, and
deeper surveys are proposed for the future,
e.g. PANSTARRS\footnote{http://pan-starrs.ifa.hawaii.edu/},
LSST\footnote{http://www.lsst.org/}, SNAP\footnote{http://snap.lbl.gov/} and others.
These surveys offer the possibility of photometric
redshifts as crude redshift estimators. With the SDSS providing a
large sample of relatively nearby galaxies, and the Hubble Deep 
Fields\footnote{http://www.stsci.edu/ftp/science/hdf/hdf.html} and the
GOODS\footnote{http://www.stsci.edu/science/goods/} survey providing
deep samples of galaxies
beyond $z=1$, many multi-color samples of galaxies have been studied
and used to estimate photometric redshifts. With good photometry 
in 4--5 optical filters, it is expected that a statistical 
accuracy characterized by an rms of $\sigma_z\simeq 0.02-0.04$ 
in $1+z$ is achievable for galaxies below $z=1$. For special 
samples such as the Large Red Galaxy (LRG) sample of the SDSS, 
one can do significantly better. Similarly it is expected that with more
filters and experience with estimating photo-$z$'s, the accuracy 
will improve and extend to higher redshifts. This is an area of
ongoing study (e.g. \citealt{ConnollyEtAl2002, 
BudavariEtAl2003, MobasherEtAl2004}). 

The accuracy of photometric redshifts determines the bin width
in redshift within which angular power spectra can be measured 
and regarded as being independent of neighboring bins (i.e. the
overlap in the galaxy distribution between neighboring bins is
small). This is important because wide bins would cause the baryon
wiggle features to be smeared out. Following \cite{SeoEisenstein2003}
we will assume that $\sigma_z< 0.04$ in $1+z$. Note that this is not a
very stringent requirement; at $z=1$, it means the rms error in
the photometric redshift is below $0.08$. 
Given a large number $N$ of galaxies with photo-$z$'s,
the mean redshift is measured accurately since the error 
in it is suppressed by $\sqrt{N}$,
which can be very small even per redshift bin for surveys of 
several 100 or 1000 square degrees. However, when the photo-$z$'s are
estimated in terms of a probability distribution over redshift per
galaxy, often the result is bi-modal or worse. Thus there are
sometimes large errors in the estimated photo-$z$, and for certain
redshift ranges they lead to systematic biases. While calibration
with spectroscopic redshifts of some fraction of the galaxies can be
used to limit such a bias, we will explore the sensitivity of our
dark-energy constraints to possible biases in the mean bin redshift. 

Our focus will be on the question: can subtle effects in the biasing
of galaxies compromise the dark-energy constraints obtained from
them? We will use the bispectrum, the Fourier transform of the 
three-point function, as an independent measure of the biasing
of galaxies. The idea of using the bispectrum in addition to the
power spectrum on large scales to
constrain both the amplitude of the mass power spectrum and bias
has been suggested and implemented (\citealt{Fry1994, Frieman1999,
  FeldmanEtAl2001, VerdeEtAl2002}). In this study we will examine
whether it can help constrain a possibly redshift dependent bias that
could mimic effects of the cosmological model we are trying to 
constrain. We will also compute the gain in signal to noise that
the bispectrum provides. We will use perturbation theory, valid
in the weakly nonlinear regime of clustering. 
\section{Theory} \label{sec:Theory}
\subsection{Dark-Energy Cosmology}
The expansion history of the universe is given by the scale factor
$a(t)$ in a homogeneous and isotropic universe.
For cosmological purposes, dark energy is associated with a 
density field, possibly weakly evolving over cosmic time, 
that affects the expansion rate of the universe, $H(a)$.
The expansion rate at late cosmic times depends
on contributions from the density of non-relativistic matter,
$\Omega_\rmn{m}$ (cold dark matter plus baryons), and the dark-energy
density, $\Omega_\rmn{de}$. Our notation is such that
$\Omega_\rmn{m}$ and $\Omega_\rmn{de}$ denote the present day values
in units of the critical density, $3H_0^2 / (8 \pi G)$. The
dark-energy formalism introduces a new, time-dependent density
described by the dark-energy equation of state:
\begin{equation}
  w(a) := \frac{p_\rmn{de}}{\rho_\rmn{de}}
        = - \frac{1}{3} \frac{\ud \ln \rho_\rmn{de}}{\ud \ln a} - 1.
\end{equation}
This density component is allowed to have an arbitrary time
dependence; it must be constrained by theory and observation. Due to
lack of compelling theoretical models for the dark energy, it is
typical to introduce a parameterization for $w(a)$ to be determined
empirically. 
We follow \cite{Linder2003} and parameterize the equation of state as
\begin{equation}
  w(a) = w_0 + w_a(1 - a).
  \label{eq:eos}
\end{equation}
Note that setting $w = -1$, i.e., $w_0 = -1$ and $w_a = 0$, corresponds
to the equation of state for a cosmological constant. 

We assume a flat universe throughout this paper. The expansion rate
for a flat universe with dark energy is given by
\begin{equation}
  H^2(a) =
    H_0^2 \left\{ \Omega_\rmn{m} a^{-3} 
      + \Omega_\rmn{de} e^{ -3 \int_1^a \ud\ln a' \left[ 1 +
                   w(a') \right] }
           \right\}.
  \label{eq:hubble}
\end{equation}
The modification in the expansion rate of the universe by a
dynamically evolving dark-energy component, as opposed to a
cosmological constant for example, alters distance measurements 
on cosmological
scales. The (comoving) distance, $\chi(a)$, from an observer at
redshift zero to a source at scale factor $a = 1 / (1 + z)$ is given
by:
\begin{equation}
  \chi(a) = \int_a^1 \frac{\ud a'}{H(a')a'^2}.
\end{equation}
Hence, given a cosmological standard ruler, such as the length scale
associated with one of the peaks of baryon-induced oscillations in the
power spectrum at a given redshift, one may obtain dark-energy
constraints via constraints on the distance function.

Dark energy that has negative pressure, i.e.,
$\rho_\rmn{de} + 3 p_\rmn{de} < 0$ as observed today, leads to
repulsive gravity and therefore does not cluster significantly. True, we
are interested in clustering on large scales in this paper, yet
spatial fluctuations in the dark energy are not expected to be
significant below the present horizon scale (\citealt{CaldwellEtAl1999, 
MaEtAl1999, Hu2002}), which we do
not approach. We thus treat dark energy as a function only of epoch.

However, since the expansion rate is altered, time-dependent dark energy
does modify the redshift evolution of mass clustering. In linear
theory, all Fourier modes of the mass density perturbation,
$\delta := \delta \rho_\rmn{m} / \bar\rho_\rmn{m}$ (relative to the
average $\bar\rho_\rmn{m}$), grow independently and
at the same rate, so
$\delta_{ \vec{k} }(a) =: D(a) \delta_{ \vec{k} }(a = 1)$
defines the growth factor, $D(a)$. The growth factor is obtained by
solving the linearized perturbation equation:
$\ddot\delta_{ \vec{k} } + 2 H \dot\delta_{ \vec{k} } = 4
\pi G \bar\rho_\rmn{m}\delta_{ \vec{k} }$, where the dot denotes
derivative with respect to physical time. This equation can be
manipulated to obtain a differential equation for the growth factor
or, alternatively, the growth rate relative to that for a flat,
matter-dominated universe, $g(a) := D(a) / a$
(\citealt{WangSteinhardt1998}, \citealt{LinderJenkins2003}):
\begin{equation}
  \frac{\ud^2 g}{\ud a^2}
    + \frac{1}{2}\left[7 - 3 w(a) \Omega_\rmn{de}(a) \right]
      \frac{1}{a} \frac{\ud g}{\ud a}
    = \frac{3}{2} \left[ w(a) - 1 \right] \Omega_\rmn{de}(a) \frac{g}{a^2}.
\end{equation}
The dark-energy density parameter at epoch $a$ is denoted
$\Omega_\rmn{de}(a)$, and is obtained from the relations
\begin{equation}
  \Omega_\rmn{de}(a) = \frac{1}{1 + X(a)}, \quad \textrm{and}
\end{equation}
\begin{equation}
  X(a) = \frac{\Omega_\rmn{m}}{ 1 - \Omega_\rmn{m} }
         e ^{-3 \int_a^1 \ud\ln a' w(a')}.
\end{equation}

We restrict our attention to the quasi-linear regime and expand
the matter density contrast perturbatively as
(e.g., \citealt{BernardeauEtAl2002}):
\begin{equation}
  \delta = D \delta_1 + D^2 \delta_2 + O(\delta^3).
\end{equation}

We suppose a biasing prescription to obtain the galaxy density
contrast from the matter density:
\begin{eqnarray}
  \delta_\rmn{g} & := & \frac{ \delta n_\rmn{g} }{ \bar n_\rmn{g} }\\
    & = & b_1 \delta + \frac{1}{2} b_2 \delta^2 + O(\delta^3).
  \label{eq:bias}
\end{eqnarray}

We allow for primordial tilt and run in the linear power spectrum
by introducing the spectral index,
$n_\rmn{s}$, and run, $\alpha := \ud \ln n_\rmn{s} / \ud \ln k$,
parameters defined at $k_0 = 0.05 / \rmn{Mpc}$, as in
\citet{KosowskyTurner1995}:
\begin{equation}
  P_\rmn{L}(k) = A \left( \frac{k}{k_0} \right)^{n_\rmn{s} + \frac{1}{2}
    \alpha \ln(k / k_0)} T^2(k).
  \label{eq:linearP}
\end{equation}
For $T(k)$, we use the transfer function of \citet{EisensteinHu1998}
that includes baryon oscillations. We determine $A$ by CMB
normalization using the relation
\begin{equation}
  A = \frac{4}{25}
      \delta_\zeta^2
      \left( \frac{H_0}{k_0} \right)^{-4}
      \left[ \frac{a_\star}{D(z_\star)} \right]^2
      \Omega_\rmn{m}^{-2}
      \left( \frac{k_0^3}{2 \pi^2} \right)^{-1}.
\end{equation}
That is, the normalization is determined by normalizing the density
perturbation amplitude at $k_0 = 0.05 / \rmn{Mpc}$ from the primordial
curvature fluctuation contrast, $\delta_\zeta$. The epoch $z_\star$
should be representative of the matter-dominated regime. The
normalization is quite insensitive to this choice. We take
$z_\star = 1088$. We will employ the CMB scalar fluctuation amplitude
parameter, $A_\rmn{s}$, whose relation to $\delta_\zeta$ is
\begin{equation}
  A_\rmn{s} = (1.84 \delta_\zeta \times 10^4)^2.
\end{equation}
\subsection{Galaxy Statistics}
We calculate the projected galaxy power spectrum using the Limber
approximation:
\begin{equation}
  P(l) =
    \frac{1}{\bar n_\rmn{g}^2}
    \int \ud\chi \, p_\rmn{g}^2(\chi)
    \frac{1}{\chi^2}
    P_\rmn{g} \left( \frac{l}{\chi}, \ a \right),
    \label{eq:P_proj}
\end{equation}
where $a = 1 / (1 + z)$ is the scale factor,
\begin{equation}
  P_\rmn{g}(k, a) = b_1^2 D^2(a) P_\rmn{L}(k), 
\end{equation}
and $D(a)$ is growth factor of the previous section.

\begin{figure}
  \centering
  \includegraphics{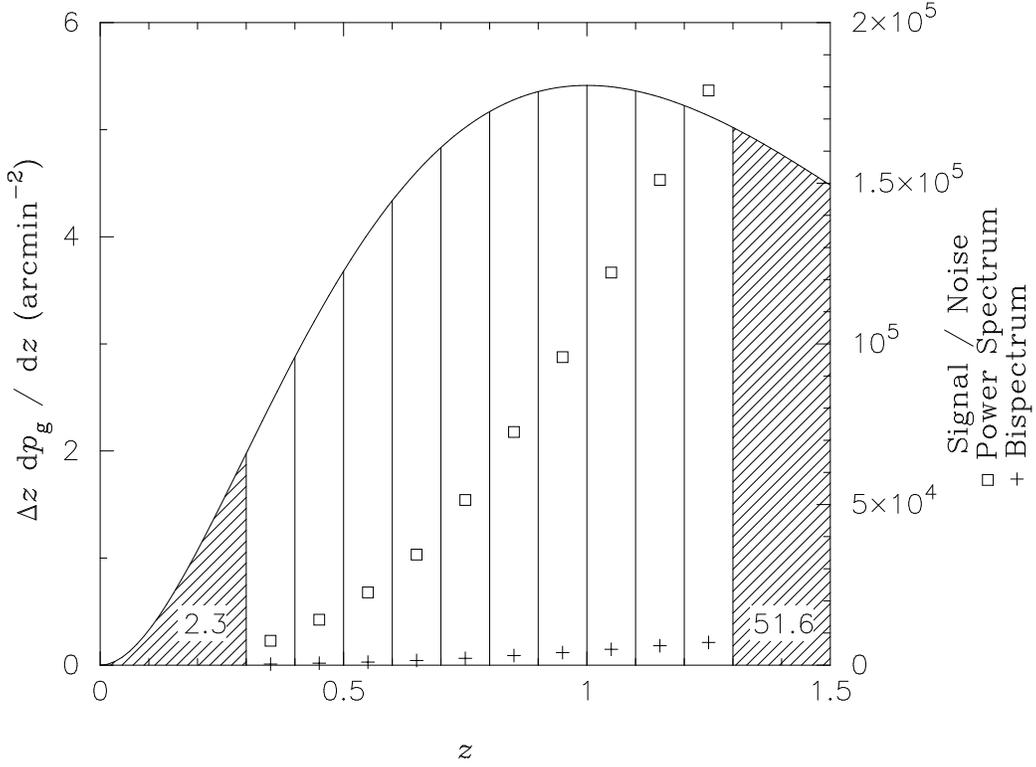}
  \caption{Galaxy redshift distribution and binning scheme employed in
    our analysis. The $y$-axis gives the number density of galaxies per 
    redshift bin of width $\Delta z=0.1$. 
    The numbers in the shaded regions
    indicate the unused galaxy number density in the low-$z$ and
    high-$z$ regions. The square symbols show the signal-to-noise 
    for the power spectrum and the crosses for the bispectrum (see
    right side of the $y$-axis)}
  \label{fig:galaxyDist}
\end{figure}

For the distribution of sample galaxies, $p_\rmn{g}(\chi)$, we use the
source distribution of \citet{Huterer2002}:
\begin{equation}
  p_\rmn{g}(z) = \bar n_\rmn{g} \frac{z^2}{2 z_0^3} e^{-z / z_0},
\end{equation}
with $z_0 = 0.5$. We separate this distribution into 10 bins over the
redshift range from $z$ = 0.3 to 1.3. All bins have equal width (0.1
in redshift), as indicated in Fig.~\ref{fig:galaxyDist}. We project
the galaxy distribution for each bin of galaxies separately using
Eq.~\ref{eq:P_proj}. A more rigorous treatment would convolve the galaxy
distribution with an appropriate selection function for a galaxy
catalog with photo-$z$ information. Our approach is a bit more crude, but we
attempt to account for redshift uncertainties by adding a nuisance
parameter for the mean redshift of each bin to our Fisher matrix, as
described in the next section.
 
Since galaxy number is conserved, we may write
\begin{equation}
  p_\rmn{g}(\chi) \, \ud\chi = p_\rmn{g}(z) \, \ud z.
\end{equation}
Finally, we rewrite Eq.~\ref{eq:P_proj} as
\begin{equation}
  P^i(l) =
    \frac{1}{(\bar n_\rmn{g}^i)^2}
    \int_i \ud \chi \, \left[ p_\rmn{g}(z) \frac{\ud z}{\ud \chi} \right]^2
    \frac{1}{\chi^2}
    P_\rmn{g} \left( \frac{l}{\chi}, \ a \right),
    \label{eq:P_g}
\end{equation}
where we have introduced the index $i$ to label the different redshift bins.

We calculate the full-sky bispectrum from the flat-sky bispectrum as
\begin{equation}
  B_{l_1 l_2 l_3} \approx
    \left( \begin{array}{ccc}
             l_1 & l_2 & l_3 \\
             0 &   0 &   0
           \end{array} \right)
    \sqrt{ \frac{ (2l_1 + 1)(2l_2 + 1)(2l_3 + 1) }{4\pi} }
    B(l_1, l_2, l_3),
\end{equation}
with the flat-sky bispectrum likewise given by a Limber-style projection:
\begin{equation}
  B^i(l_1, l_2, l_3) =
    \frac{1}{(\bar n_\rmn{g}^i)^3} 
    \int_i \ud\chi \, \left[ p_\rmn{g}(z) \frac{\ud z}{\ud\chi} \right]^3
    \frac{1}{\chi^4}
    B_\rmn{g}\left( \frac{l_1}{\chi}, \frac{l_2}{\chi},
      \frac{l_3}{\chi}, \ a \right).
    \label{eq:B_g}
\end{equation}
To $O(\delta^4)$, the galaxy bispectrum is given by the sum of two
terms:
\begin{eqnarray}
  B_\rmn{g}(k_1, k_2, k_3, a) & = &
    b_1^3 D^4(a) \left[ 2 F_2(k_1, k_2) P(k_1) P(k_2)
                + \rmn{cyclic} \ k_i \ \rmn{permutations} \right]
                \nonumber \\
    & + & \frac{1}{2}b_1^2 b_2 D^4(a) \left[ 2 P(k_1) P(k_2)
                + \rmn{cyclic} \ k_i \ \rmn{permutations} \right],
\end{eqnarray}
where
\begin{equation}
  F_2(k_1, k_2) =
    \frac{5}{7} 
    + \frac{1}{2} \frac{\vec k_1 \cdot \vec k_2}{k_1 k_2}
      \left( \frac{k_1}{k_2} + \frac{k_2}{k_1} \right)
    + \frac{2}{7} \left(\frac{\vec k_1 \cdot \vec k_2}{k_1 k_2}\right)^2
\end{equation}
is the second-order perturbation theory kernel. Note that the two
terms have different triangle configuration dependence, and different
dependence on $b_1$ and $b_2$.
\subsection{Covariance and Fisher Matrix Formalism}
The power spectrum covariance for redshift bin $i$ ignores the contribution from the
trispectrum, which should be small in the linear regime:
\begin{equation}
  C_{l, l'}^i = \frac{1}{f_\rmn{sky}} \frac{2}{2l + 1} \left[ P^i(l) + \frac{1}{\bar n_\rmn{g}^i}
  \right]^2 \delta_{l l'}.
\end{equation}
Here $f_\rmn{sky}$ is the fraction of sky coverage for the survey,
for which we use the value 0.1. $\bar n_\rmn{g}^i$ is the galaxy density for redshift bin
$i$. In our galaxy distribution (Fig.~\ref{fig:galaxyDist}) it is 2.4 galaxies per arcmin${}^2$ for the lowest redshift
bin, and 5.4 galaxies per arcmin${}^2$ at the peak of the galaxy distribution.

The bispectrum covariance ignores contributions from 3-,
4-, and 6-point functions, which are also expected to be small:
\begin{equation} \begin{split}
  C_{l_1 l_2 l_3, l'_1 l'_2 l'_3}^i = &
    \frac{1}{f_\rmn{sky}}
    \left[ P(l_1) + \frac{1}{\bar n_\rmn{g}^i} \right]
    \left[ P(l_2) + \frac{1}{\bar n_\rmn{g}^i} \right]
    \left[ P(l_3) + \frac{1}{\bar n_\rmn{g}^i} \right] \\
     & \times \big(  \delta_{l_1, l'_1} \delta_{l_2, l'_2} \delta_{l_3, l'_3}
        + \delta_{l_1, l'_1} \delta_{l_2, l'_3} \delta_{l_3, l'_2}
        + \delta_{l_1, l'_2} \delta_{l_2, l'_1} \delta_{l_3,l'_3}
    +  \delta_{l_1, l'_2} \delta_{l_2, l'_3} \delta_{l_3, l'_1}
        + \delta_{l_1, l'_3} \delta_{l_2, l'_1} \delta_{l_3, l'_2}
        + \delta_{l_1, l'_3} \delta_{l_2, l'_2} \delta_{l_3, l'_1}
     \big).
\end{split} \end{equation}

We assume that the likelihood function for the galaxy power spectrum
and galaxy bispectrum is Gaussian, and employ a Fisher matrix analysis
to approximate the likelihood near the fiducial cosmological model
defined by the parameters in Table \ref{tab:params}.
\begin{table}
  \begin{tabular}{rlll}
    \hline
    Parameter & Fiducial Value & Description & Calculated As\\
    \hline
    $\omega_\rmn{b}$ & 0.023 & Baryon Physical Density\\
    $\omega_\rmn{d}$ & 0.112 & Dark-Matter Physical Density\\
    $\Omega_\rmn{de}$ & 0.69 
                      & Dark-Energy Density Parameter (Eq.~\ref{eq:hubble})\\
    $w_0$ & -1 & Equation of State Parameter (Eq.~\ref{eq:eos})\\
    $w_a$ & 0 & Equation of State Parameter (Eq.~\ref{eq:eos})\\
    $A_\rmn{s}$ & 0.82 
                & Scalar Fluctuation Amplitude at $k = 0.05 / \rmn{Mpc}$\\
    $n_\rmn{s}$ & 0.979 
                & Primordial Spectral Index at $k = 0.05 / \rmn{Mpc}$ (Eq.~\ref{eq:linearP})\\
    $\alpha$ & 0 & Primordial Run ($= d \ln n_\rmn{s} / d \ln k$) at
                   $k = 0.05 / \rmn{Mpc}$\\
    $b_1$ & 0.998 & First Order Galaxy Bias Factor (Eq.~\ref{eq:bias})\\
    $b_2$ & 0 & Second Order Galaxy Bias Factor (Eq.~\ref{eq:bias})\\
    $\tau$ & 0.143 & Optical Depth\\
    \hline
    $\Omega_\rmn{m}$ & 0.31 & Matter Density
                     & $(\omega_\rmn{b} + \omega_\rmn{d}) / h^2$\\
    $\Omega_\rmn{b}$ & 0.053 & Baryon Density
                     & $\omega_\rmn{b} / h^2$\\
    $\Omega_\rmn{tot}$ & 1 & Total Density (Assume Flat Cosmology)\\
    $h$ & 0.66 & Current Hubble Parameter in Units of 100 km / s / Mpc
        & $\sqrt{( \omega_\rmn{b} + \omega_\rmn{d} ) 
           / ( 1 - \Omega_\rmn{de} )}$\\
    $\sigma_8$ & 0.88 & Galaxy-Scale Fluctuation Amplitude\\
    \hline
  \end{tabular}
  \caption{The fiducial cosmological model used throughout this paper.}
  \label{tab:params}
\end{table}

The Fisher matrix elements are given by
\begin{equation}
  F_{\alpha \beta}^P =
    \sum_i
    \sum_{ l_\rmn{min} \le l \le l_\rmn{max} }
      \frac{ \partial P(l) }{\partial p_\alpha}
      (C_{l, l}^i)^{-1}
      \frac{ \partial P(l) }{\partial p_\beta}
      \label{eq:FP}
\end{equation}
for the power spectrum, and
\begin{equation}
  F_{\alpha \beta}^B =
    \sum_i
    \sum_{ l_\rmn{min} \le l_1 \le l_2 \le l_3 \le l_\rmn{max} }
      \frac{ \partial B_{l_1 l_2 l_3} }{\partial p_\alpha}
      (C_{l_1 l_2 l_3, l_1 l_2 l_3}^i)^{-1}
      \frac{ \partial B_{l_1 l_2 l_3} }{\partial p_\beta}
      \label{eq:FB}
\end{equation}
for the bispectrum. The derivatives are evaluated at the fiducial model.

The large-scale limit, $l_\rmn{min}$, is introduced because the Limber
approximation fails when probing scales
larger than the line-of-sight width of the given bin of
galaxies. So we let $l_\rmn{min}$ be the wavenumber that
corresponds to the width of the bin. Since the information 
from baryon oscillations is contained at significantly larger $l$
in any case, this choice has little impact on our results. 

\begin{figure}
  \centering
  \includegraphics{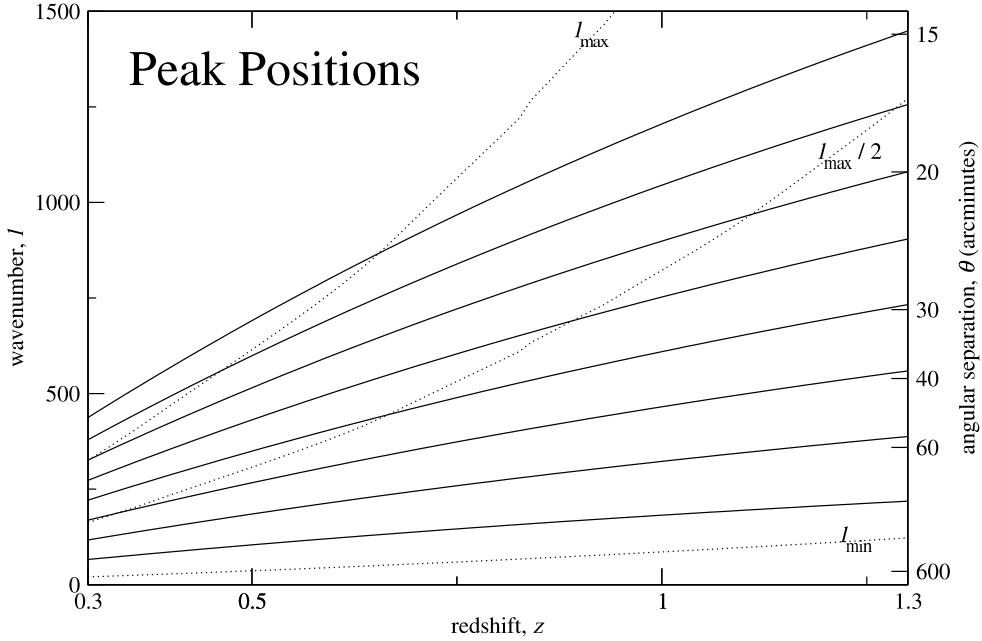}
  \caption{Angular separation and wavenumber of the first five peaks
    in the galaxy power spectrum induced by
    baryon oscillations as a function of galaxy redshift. The dotted
    lines correspond to our cutoff values $l_\rmn{max}$ and
    $l_\rmn{min}$. The $l(\theta)$ relation is approximate---we use
    $l = 2\pi/\theta$.}
  \label{fig:peaks}
\end{figure}

The small-scale limit, $l_\rmn{max}$, is meant to represent the scale
beyond which perturbation theory is unacceptable. Nonlinear effects
tend to smooth out the high-$l$ baryon oscillations in any case
(\citealt*{MeiksinWhitePeacock1999}). We define this
(redshift-dependent) scale as $l_\rmn{max} = k_\rmn{max} / \chi(z)$, where
$k_\rmn{max}$ is defined by
\begin{equation}
  \sigma^2 = \int_0^{ k_\rmn{max} } \ud^3 k \, P(k) = 1.
\end{equation}
The values of $l_\rmn{max}$ are plotted as the upper dotted line in
Fig.~\ref{fig:peaks}. We will test below the sensitivity of our
results to the choice of $l_\rmn{max}$  by repeating our calculations
for $l_\rmn{max}/2$. 
 
We neglect the cross covariance between the power spectrum and
bispectrum, which is proportional to the 5-point function, so that the
Fisher information matrix for the power spectrum and bispectrum
combined is simply the sum
\begin{equation}
  F_{\alpha \beta} \approx F_{\alpha \beta}^P + F_{\alpha \beta}^B.
\end{equation}
Finally, for a pair of parameters $p_\alpha$ and $p_\beta$, we
calculate $\chi^2$ as
\begin{equation}
  \chi_{p_\alpha, p_\beta}^2 =
    \left( \begin{array}{c}
           \Delta p_\alpha \\
              \Delta p_\beta
           \end{array}
    \right)
    \left[ \begin{array}{cc}
           \left( \mathbf{F}^{-1} \right)_{\alpha \alpha} & 
              \left( \mathbf{F}^{-1} \right)_{\alpha \beta} \\
           \left( \mathbf{F}^{-1} \right)_{\beta \alpha} &
              \left( \mathbf{F}^{-1} \right)_{\beta \beta}
            \end{array}
    \right]^{-1}
    \left( \begin{array}{c}
           \Delta p_\alpha \\
              \Delta p_\beta
           \end{array}
    \right),
\end{equation}
and identify the one-sigma contour as that contour defined by $\chi^2 =
2.3$.

For a flux-limited sample of galaxies, one should allow for a 
dependence of the galaxy bias on redshift. Hence we include an
independent $b_1^i$ and $b_2^i$ for each redshift bin in our Fisher
matrix.

We allow for biases in photometric redshifts by 
including the mean redshift of
each galaxy bin as a nuisance parameter in the Fisher matrix. We 
set the rms in the mean redshift to be $0.01$ in $1+z$
for each redshift bin. 
This is conservative, since samples of spectroscopic
redshifts will likely be available for calibration and the residual 
bias should be significantly smaller. 
\section{Results} \label{sec:results}
\subsection{Small Parameter Set}
Since we are most interested in constraining the dark energy and
galaxy bias, we first consider a restricted set of cosmological
parameters: the dark-energy parameters, the overall
amplitude normalization, and the bias parameters:
$\{w_0, w_a, A_s, b_1, b_2\}$. We will impose no external priors on
these parameters, but all other parameters are fixed to the fiducial
values in Table \ref{tab:params} with the exception of
$\Omega_\rmn{de}$, which we marginalize over after applying a prior of
0.05. It is important to constrain $\Omega_\rmn{de}$ to be able
to extract information about the other parameters. Fortunately, CMB
information can be used to constrain $\Omega_\rmn{de}$ well.
In the next subsection, we use a more complete set of parameters and
combine our results with information from the CMB, and further allow for a
possibly redshift-dependent galaxy bias by giving each redshift bin an
independent $b_1^i$ and $b_2^i$.

\begin{figure}
  \centering
  \ifthenelse{\boolean{grey-scale}}
             {\includegraphics{PModels_grey}
              \includegraphics{PModelsNW_grey}}
             {\includegraphics{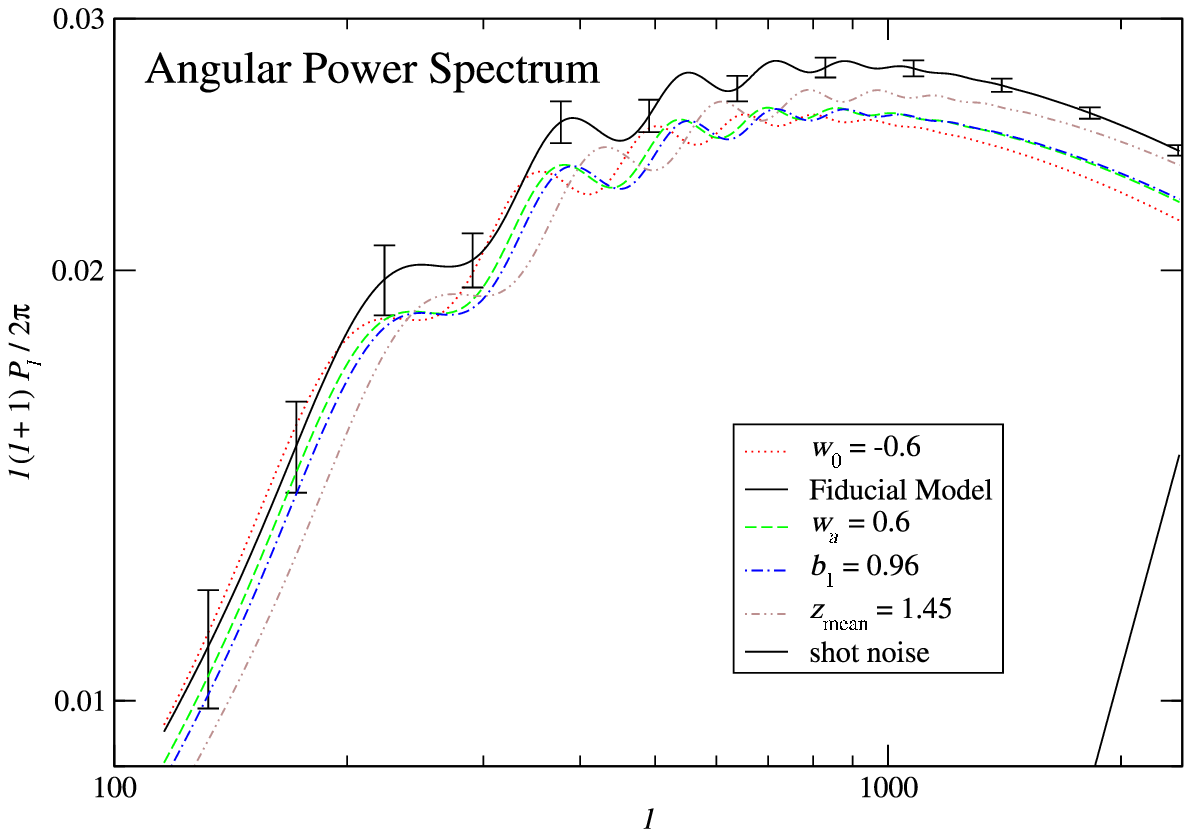}
              \includegraphics{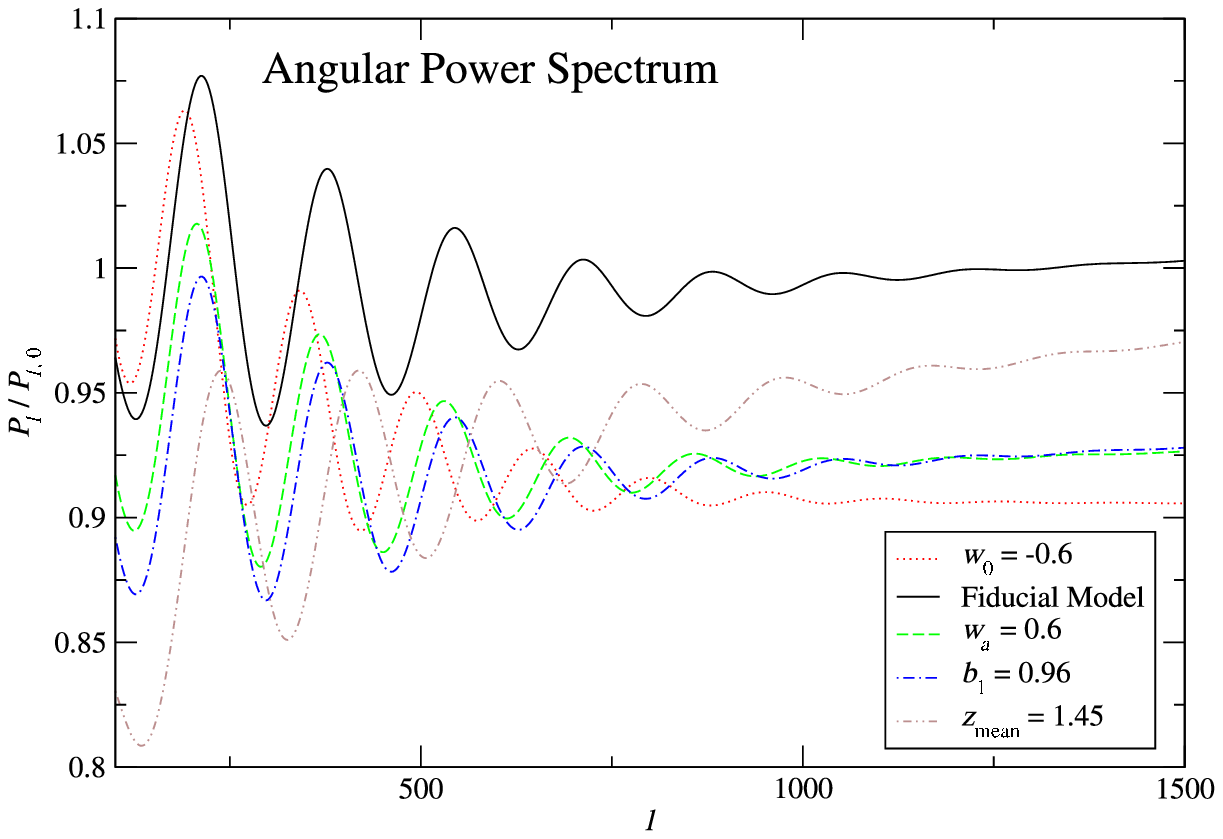}}
  \caption{Effects of dark energy and other parameter variations on
    the projected galaxy power spectrum
    (Eq.~\ref{eq:P_g}). Except for the one labelled exception, we plot
    the projected power spectrum for our redshift bin with
    $z_\rmn{mean} = 1.25$. In the lower panel, we divide by a smooth
    spectrum corresponding to zero baryon density. This shows the 
    differences between the models more clearly. Note that the 
    shot noise curve for the lower redshift bins will be more 
    than a factor of two larger than what is plotted here for
    $z=1.25$. Even so, it is clear that shot noise is not a significant 
    source of error for the survey parameters we have used.}
  \label{fig:Pl}
\end{figure}

\begin{figure}
  \centering
  \ifthenelse{\boolean{grey-scale}}
             {\includegraphics{Beq_grey}
              \includegraphics{Bis_grey}}
             {\includegraphics{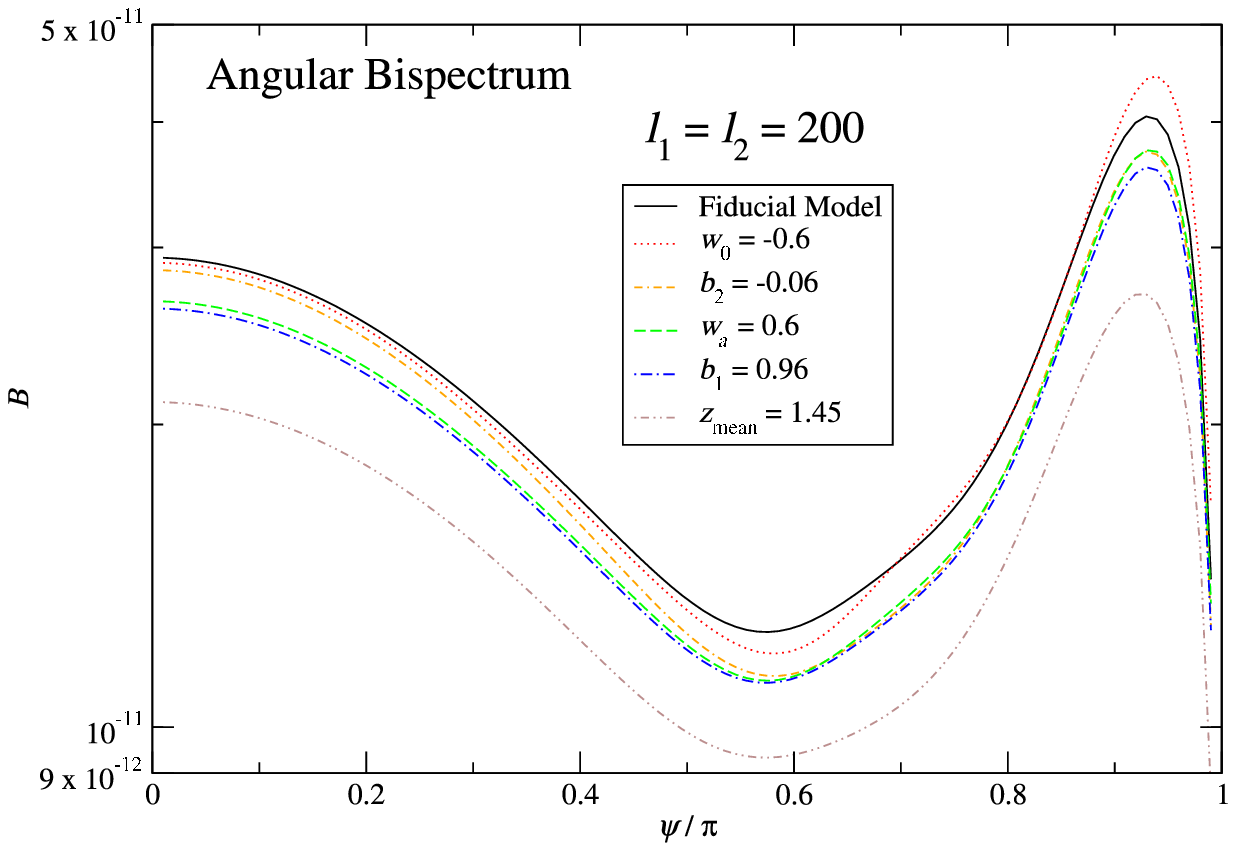}
              \includegraphics{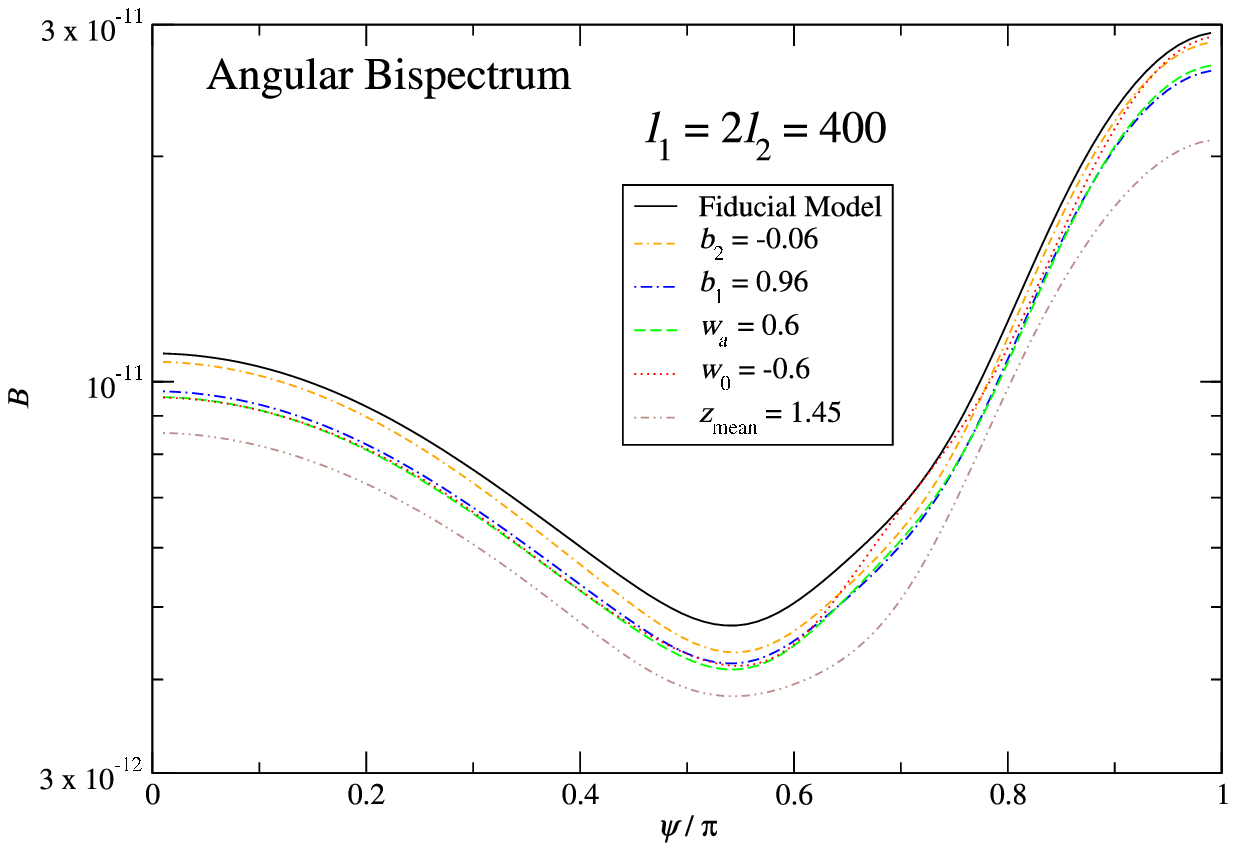}}
  \caption{Configuration dependence of the projected bispectrum for
    various parameters choices as in Fig.~\ref{fig:Pl}. The upper
    panel uses equilateral triangles, while the lower panel uses
    isosceles triangles. Note that the effect of the bias
    parameter $b_2$, which leads to nonlinearity in the biasing, on 
    the bispectrum is at the same level as that of $b_1$. Hence we 
    expect that the bispectrum can be used to constrain this parameter. } 
  \label{fig:BModels}
\end{figure}

Since the dark-energy parameters affect the angular diameter
distances, they shift the position of the peaks in the
projected galaxy power spectrum and bispectrum. In Fig.~\ref{fig:Pl},
we plot the angular galaxy power spectrum for our highest redshift
bin: $1.2 \le z \le 1.3$. Notice that, in addition to peak
shifting, the overall amplitude of the fluctuations is also changed
due to the growth factor's dependence on dark energy. 
Because of this, we include
those parameters that affect the amplitude of the galaxy statistics
in our 5-parameter model. We neglect redshift
uncertainties, and any redshift dependence of the bias
parameters. That is, for this case, a single $b_1$ and $b_2$
parameter are used for all 10 redshift bins. Also, we plot the galaxy
bispectrum signal in Fig.~\ref{fig:BModels}, which shows
how these parameters modify the configuration dependence of the bispectrum.

\begin{figure}
  \centering
  \includegraphics{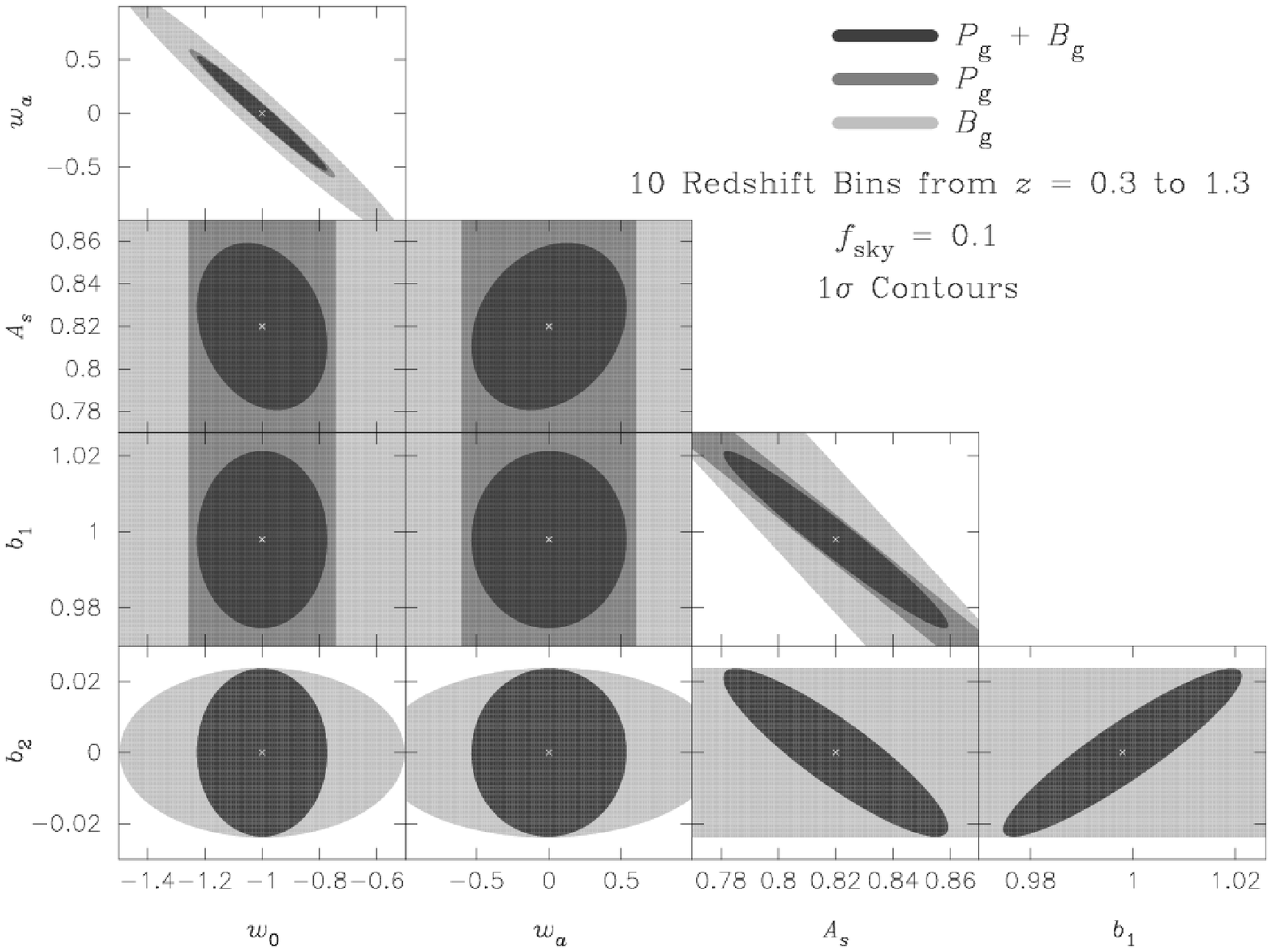}
  \caption{1-$\sigma$ constraints from the Fisher matrix analysis of
    the 5-parameter cosmological model. The fiducial model is the
    cosmological constant model in Table \ref{tab:params}, and is
    marked with an $\times$ in the figure.}
  \label{fig:5params}
\end{figure}

The results of our Fisher matrix analysis for the 5-parameter model
are presented in Fig.~\ref{fig:5params}. These are two dimensional
(marginalized) projections of 1-$\sigma$ contours in the 5-parameter
space. Based on these contours, the addition of the
bispectrum information does not appear to strengthen the dark-energy
constraints substantially.
However, constraints on the other three parameters are all
significantly improved by the bispectrum information. This is not too
surprising, for the galaxy power spectrum is proportional to $A_\rmn{s}
b_1^2$ (Eq.~\ref{eq:P_g}), and so the galaxy power spectrum alone
cannot distinguish $A_\rmn{s}$ and $b_1$. The degeneracy is broken by
the bispectrum, as its dependence is $A_\rmn{s}^2 b_1^3$
(for $b_2 = 0$, Eq.~\ref{eq:B_g}). The degeneracy breaking is most
dramatic in the $A_\rmn{s}-b_1$ plot of Fig.~\ref{fig:5params}, where
the power spectrum or bispectrum information alone results in
unbounded contours, but each contour comes with a different slopes, so
that a tight contour is obtained when the two statistics are combined.

To lowest order in perturbation theory, the power spectrum does not
depend on $b_2$. Dependence on $b_2$ enters at next order, proportional
to the trispectrum.  We have neglected the trispectrum contribution 
relative to the power spectrum, and so there are no power spectrum 
contours for $b_2$ in Fig.~\ref{eq:P_g}. However our formalism allows
for this contribution to be included; with the bispectrum we can 
actually measure $b_2$ and then use the full power spectrum up to 
second order, which thus allows for a nonlinear, scale-dependent
bias. The bispectrum enables constraints on $b_2$ because, 
as can be seen from Eq.~\ref{eq:B_g}, it contains two terms
with different triangle configuration dependence. Only the second of
the two terms
depends on $b_2$, so observation of the galaxy bispectrum is useful to
constrain this parameter.
\subsection{Large Parameter Set}
We have done the Fisher matrix analysis for a
comprehensive set of cosmological parameters: $\{\omega_\rmn{b}, 
\omega_\rmn{d},
\Omega_\rmn{de}, w_0, w_a, A_\rmn{s}, n_\rmn{s}, \alpha, b_1^i, b_2^i,
\tau \}$. Refer to Table~\ref{tab:params} for a description of each
parameter and its fiducial value. We use an independent pair of bias
parameters for each redshift bin. We also include the mean redshift
values of our redshift bins as additional nuisance parameters. We set
the rms uncertainty to be $0.01$ in $1+z$ for the mean of each
redshift bin.

\begin{figure}
  \centering
  \ifthenelse{\boolean{grey-scale}}
             {\includegraphics{PDerivatives_grey}}
             {\includegraphics{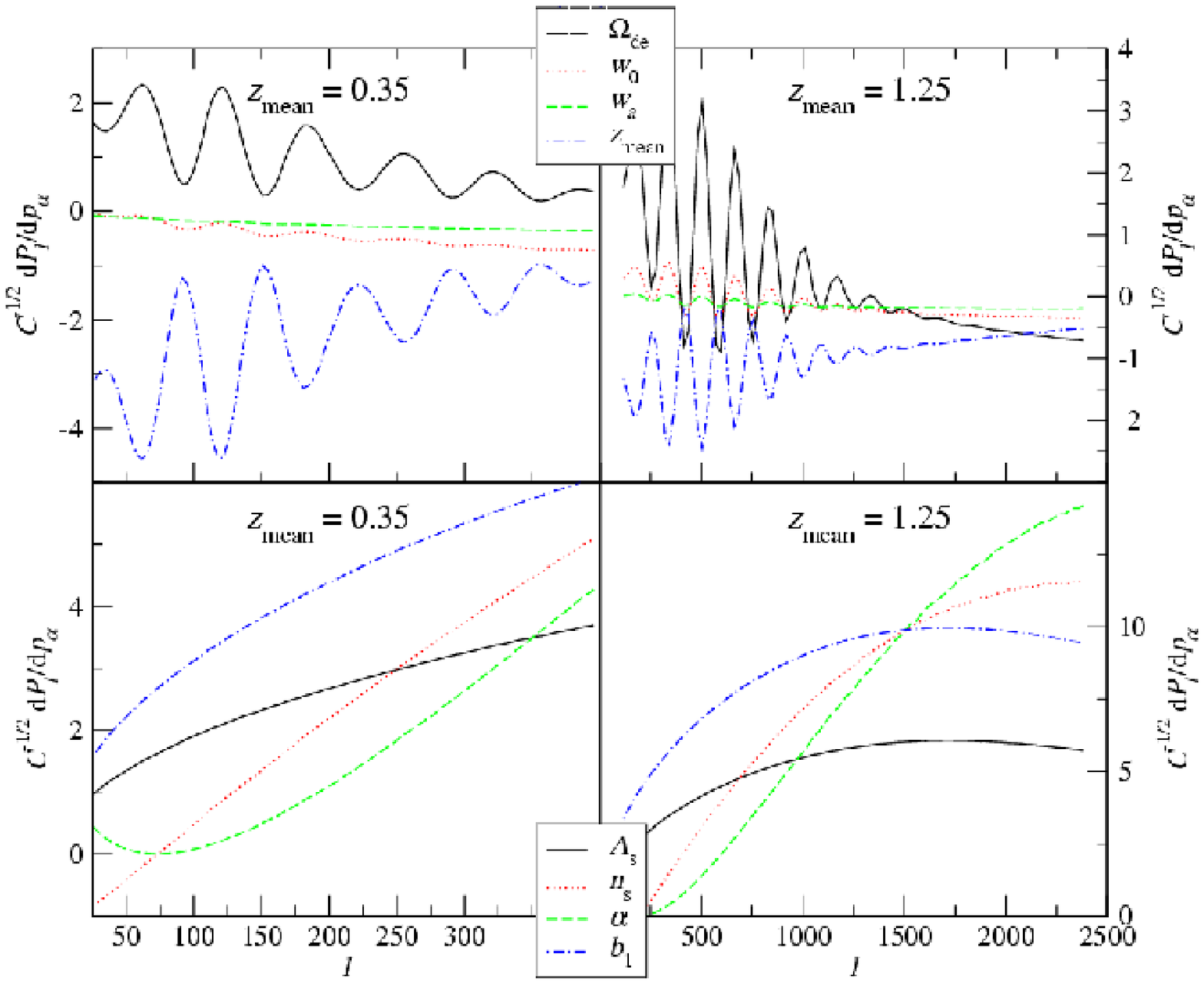}}
  \caption{Plot indicating the sensitivity of the power spectrum
    to cosmological parameters as a function of
    $l$. The curves have been weighted by the covariance $1 /
    \sqrt{C}$, as in Eq.~\ref{eq:FP}.}
  \label{fig:PDerivatives}
\end{figure}

\begin{figure}
  \centering
  \ifthenelse{\boolean{grey-scale}}
             {\includegraphics{BDerivatives_grey}}
             {\includegraphics{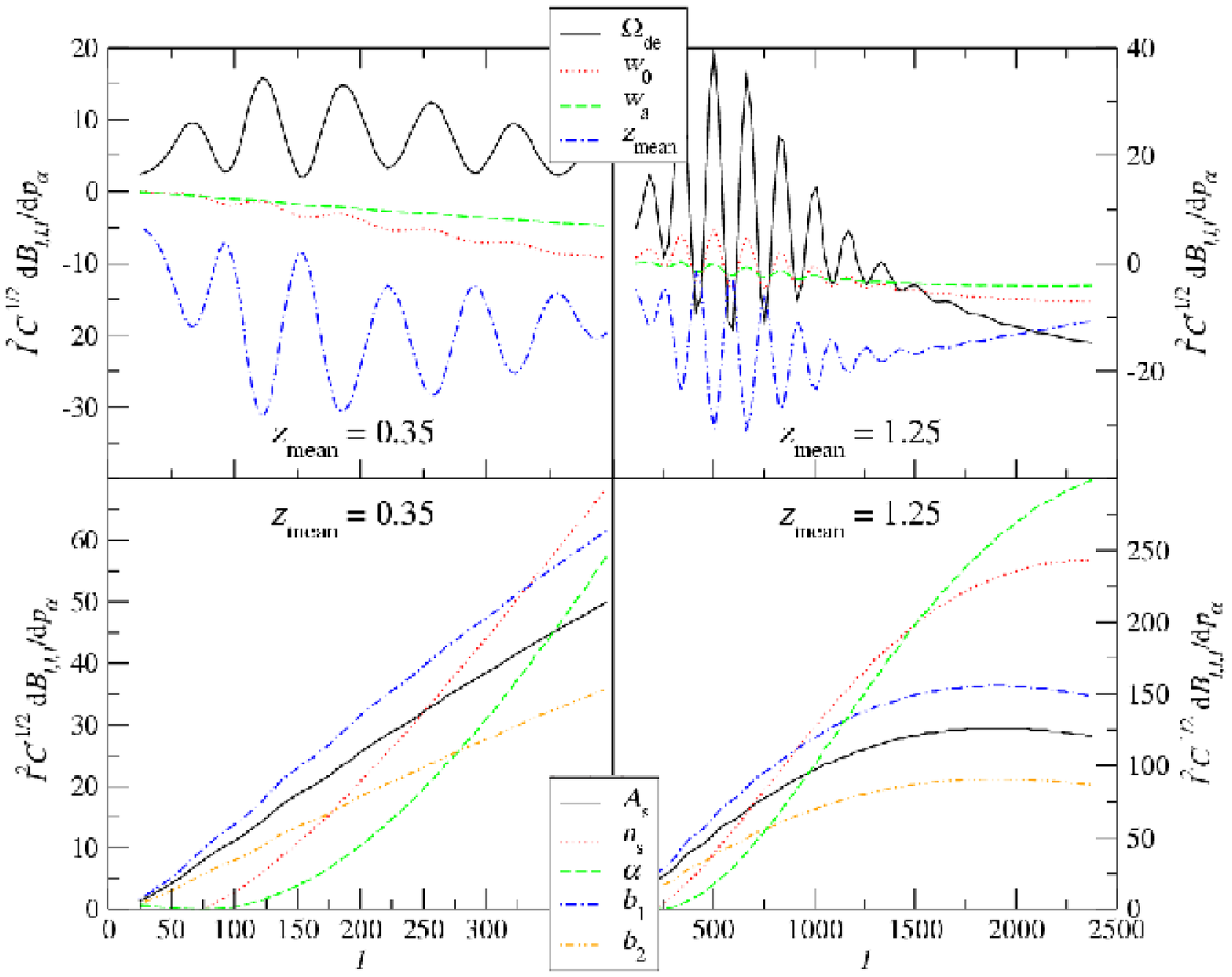}}
  \caption{Plot indicating the sensitivity of the bispectrum
    to cosmological parameters as a function of
    $l$. The curves have been weighted by the covariance $1 /
    \sqrt{C}$, as in Eq.~\ref{eq:FB}, and by a factor of
    $l^2$ as an estimate of the number of triangle configurations.}
  \label{fig:BDerivatives}
\end{figure}

To get a feel for the sensitivity of the power spectrum and bispectrum
signals to each of these parameters, we have plotted in
Fig.~\ref{fig:PDerivatives} and \ref{fig:BDerivatives} the derivatives with
respect to each parameter, evaluated at the fiducial model, as a
function of $l$. The derivatives are weighted by a covariance factor,
$1 / \sqrt{C}$, as this is the weighting that enters the Fisher
matrices (cf. Eqs.~\ref{eq:FP} and \ref{eq:FB}). Additionally, for
the bispectrum signal, a sum over all triangle configurations should
grant a proportionately higher sensitivity for larger $l$ values. We
attempt to roughly account for this by weighting the bispectrum
derivatives by a factor of
$l^2$. Thus, squaring each weighted power spectrum and bispectrum
derivative and summing is approximately (unmarginalized) $1 / \sigma^2$ for
that parameter.

\begin{figure}
  \includegraphics{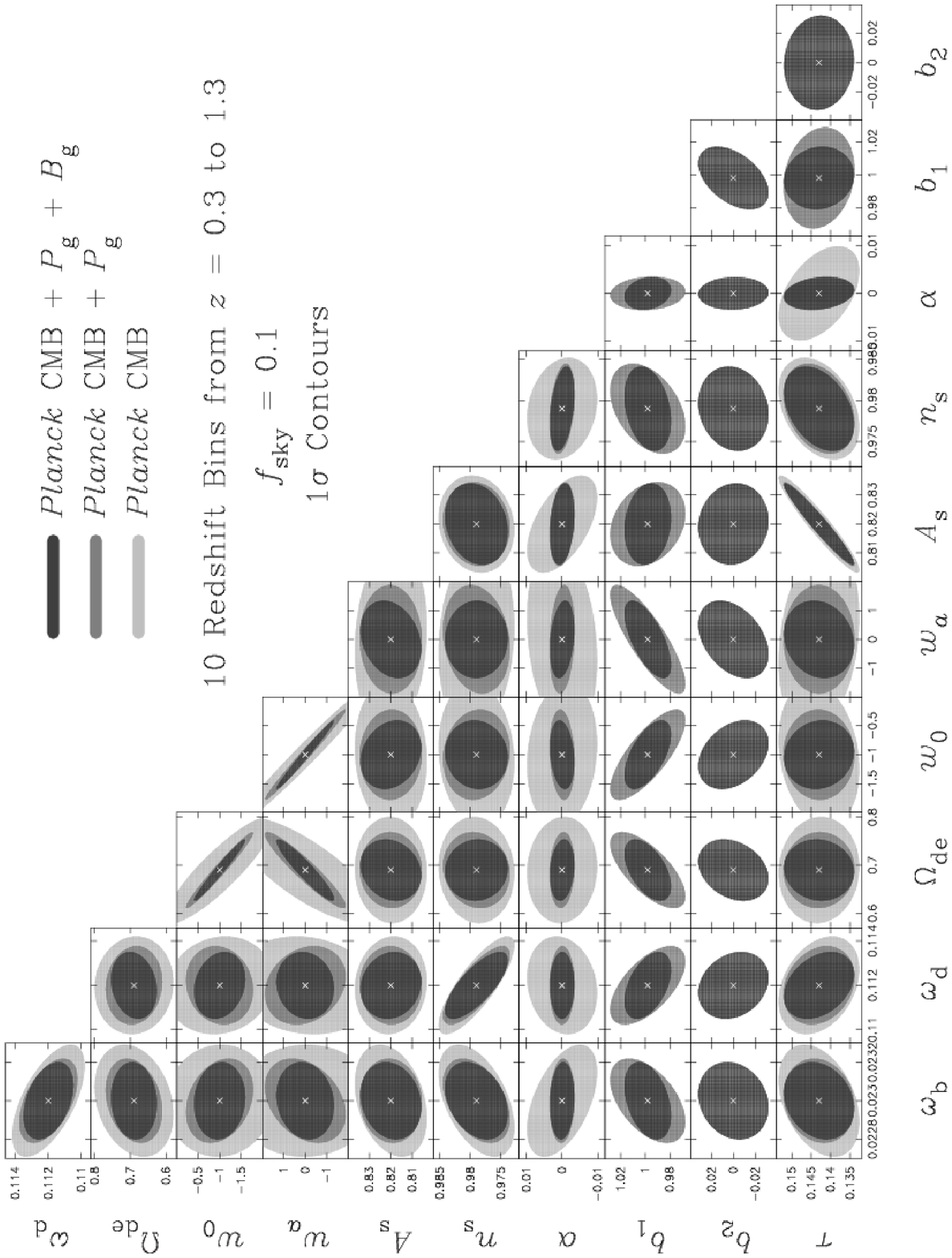}
  \caption{1-$\sigma$ constraints from the Fisher matrix analysis of
    the 11-parameter cosmological model. We have combined our
    galaxy power spectrum and bispectrum Fisher matrices with a CMB Fisher
    matrix calculated using projected \satellite{Planck}
    covariances. The fiducial model is the cosmological constant model
    in Table \ref{tab:params}, and is marked with an $\times$ in the
    figure.}
  \label{fig:11params}
\end{figure}

With so many free parameters, it is
necessary to apply some priors or to combine galaxy data with some 
complementary cosmological information, such as the CMB anisotropy. 
Galaxy information by itself suffers from strong degeneracies, like 
$w_0$-$w_a$ in particular. The marginalized one-sigma constraints 
from galaxy data
alone are weak: $w_0$ constrained to $\pm 1$ and $w_a$ constrained to $\pm
4$. However, combination with the complementary CMB information gives
interesting constraint ellipses, as shown in 
Fig.~\ref{fig:11params}, or Table \ref{tab:sigmas}. We have not 
explored other possible cosmological probes with which to combine angular
galaxy clustering. 

We plot the one-sigma constraints obtained from our Fisher matrix
analysis in Fig.~\ref{fig:11params}. 
We calculate CMB Fisher matrices in the same way as was done for
\cite{TakadaJain2004}. CMB temperature and polarization power spectra
and the cross spectrum are calculated using CMBFAST version 4.5
(\citealt{SeljakZaldarriaga1996}). We assume the experimental
specifications of the \satellite{Planck} 143 and 217 GHz channels with 65 per cent
sky coverage. We also assume a spatially flat universe with no massive
neutrinos and no gravity waves.
We summarize our constraint prospects in Table \ref{tab:sigmas}.
\begin{table}
  \begin{tabular}{c r @{.} l r @{.} l r @{.} l r @{.} l r @{.} l}
    \hline
    parameter    & \multicolumn{2}{c}{$\rmn{CMB}$} &
                   \multicolumn{2}{c}{$\rmn{CMB} + P_\rmn{g}$} &
                   \multicolumn{2}{c}{$\rmn{CMB} + P_\rmn{g} + B_\rmn{g}$} &
                   \multicolumn{2}{c}{$\rmn{CMB} + \rmn{SNe}$} &
                   \multicolumn{2}{c}{$\rmn{CMB} + \rmn{SNe} + P_\rmn{g} + B_\rmn{g}$} \\
    \hline
    $\omega_\rmn{b}$ & 0&00019 &     0&00014 &     0&00013 &     0&00018 &     0&00013\\
    $\omega_\rmn{d}$ & 0&00145 &     0&00116 &     0&00099 &     0&00141 &     0&00096\\
    $\Omega_\rmn{de}$ & 0&0727 &      0&0520 &      0&0421 &      0&0061 &      0&0044\\
    $w_0$  & 0&818 &       0&511 &       0&389 &       0&077 &       0&056\\
    $w_a$ & 2&180 &       1&233 &       0&881 &       0&305 &       0&223\\
    $A_\rmn{s}$  &     0&0111 &     0&0096 &     0&0093 &     0&0108 &     0&0091\\
    $n_\rmn{s}$ &     0&0041 &     0&0036 &     0&0034 &     0&0039 &     0&0034\\
    $\alpha$ &     0&0065 &     0&0023 &     0&0023 &     0&0059 & 0&0022\\
    $b_1$ & \multicolumn{2}{c}{} & 0&0202 & 0&0125 & \multicolumn{2}{c}{} & 0&0100\\
    $b_2$ & \multicolumn{2}{c}{} & \multicolumn{2}{c}{} & 0&021 & \multicolumn{2}{c}{} & 0&020 \\
    $\tau$ & 0&0068 &     0&0060 &     0&0059 &     0&0065 &    0&0058\\
    \hline
  \end{tabular}

  \vspace{10pt}
  \begin{tabular}{c r @{.} l r @{.} l r @{.} l r @{.} l r @{.} l}
    \hline
                 & \multicolumn{2}{c}{} &
                   \multicolumn{2}{c}{}&
                   \multicolumn{2}{c}{SUGRA} & 
                   \multicolumn{2}{c}{SUGRA} &
                   \multicolumn{2}{c}{SUGRA} \\
    parameter    & \multicolumn{2}{c}{$\rmn{CMB} + P_\rmn{g}^{z < 2.3}$} & 
                   \multicolumn{2}{c}{$\rmn{CMB} + P_\rmn{g}^{z < 2.3} + B_\rmn{g}^{z < 2.3}$} &
                   \multicolumn{2}{c}{$\rmn{CMB} + P_\rmn{g}$} &
                   \multicolumn{2}{c}{$\rmn{CMB} + P_\rmn{g} + B_\rmn{g}$} &
                   \multicolumn{2}{c}{$\rmn{CMB} + \rmn{SNe}$}\\
    \hline
    $\omega_\rmn{b}$ & 0&00013 &     0&00012 &     0&00014 &     0&00013 &     0&00019\\
    $\omega_\rmn{d}$ &     0&00094 &     0&00075 &     0&00123 &     0&00096 &     0&00150\\
    $\Omega_\rmn{de}$ &   0&0439 &      0&0361 &      0&0064 &      0&0062 &      0&0061\\
    $w_0$ &        0&400 &       0&311 &       0&020 &       0&020 &       0&017\\
    $w_a$ &   0&941 &       0&687 &       0&038 &       0&033 &       0&038\\
    $A_\rmn{s}$ &  0&0095 &     0&0089 &     0&0092 &     0&0091 &     0&0103\\
    $n_\rmn{s}$ &     0&0033 &     0&0031 &     0&0038 &     0&0034 &     0&0041\\
    $\alpha$ &    0&0018 &     0&0017 &     0&0023 &     0&0023 &     0&0061\\
    $b_1$ & 0&0165 & 0&0097 & 0&0211 & 0&0140 & \multicolumn{2}{c}{}\\
    $b_2$  &    \multicolumn{2}{c}{} &    0&020 &    \multicolumn{2}{c}{} &
                      0&024 &    \multicolumn{2}{c}{} \\
    $\tau$ &     0&0059 &     0&0057 &     0&0059 &     0&0059 &     0&0065\\
    \hline
  \end{tabular}
  \caption{Summary of marginalized 1-$\sigma$ prospects. Except where
    indicated otherwise, $P_\rmn{g}$ and $B_\rmn{g}$ include galaxies
    to redshift 1.3. The bias parameters listed here are those for the
    redshift bin with $0.9 \le z \le 1.0$.}
  \label{tab:sigmas}
\end{table}

As in Fig.~\ref{fig:5params}, the bispectrum helps primarily
in constraining the amplitude and bias parameters. This is 
important because poor constraints on the $b_1$ and $b_2$ parameters
mean that the dark-energy constraints may not be robust against 
possibly scale-dependent or nonlinear bias. The main difference
in the level of accuracy achieved compared to the 5-parameter study
is due to the bias parameters, since here we allow for two independent
parameters in each redshift bin. Constraints on the biasing from other
observations, or having some basis to reduce the parameter set from the
twenty we use here, would improve dark-energy constraints by up to 
a factor of two.

Fig.~\ref{fig:11params} illustrates the power of a wide area
multi-color survey, in that the full set of parameters gets
constrained by the galaxy power spectrum and bispectrum, in 
conjunction with the CMB. For example, the constraints on the 
shape of the primordial power spectrum and on dark-energy 
parameters are interesting enough that increasing survey area
or depth becomes appealing. We discuss these below.

\begin{figure}
  \centering
  \includegraphics{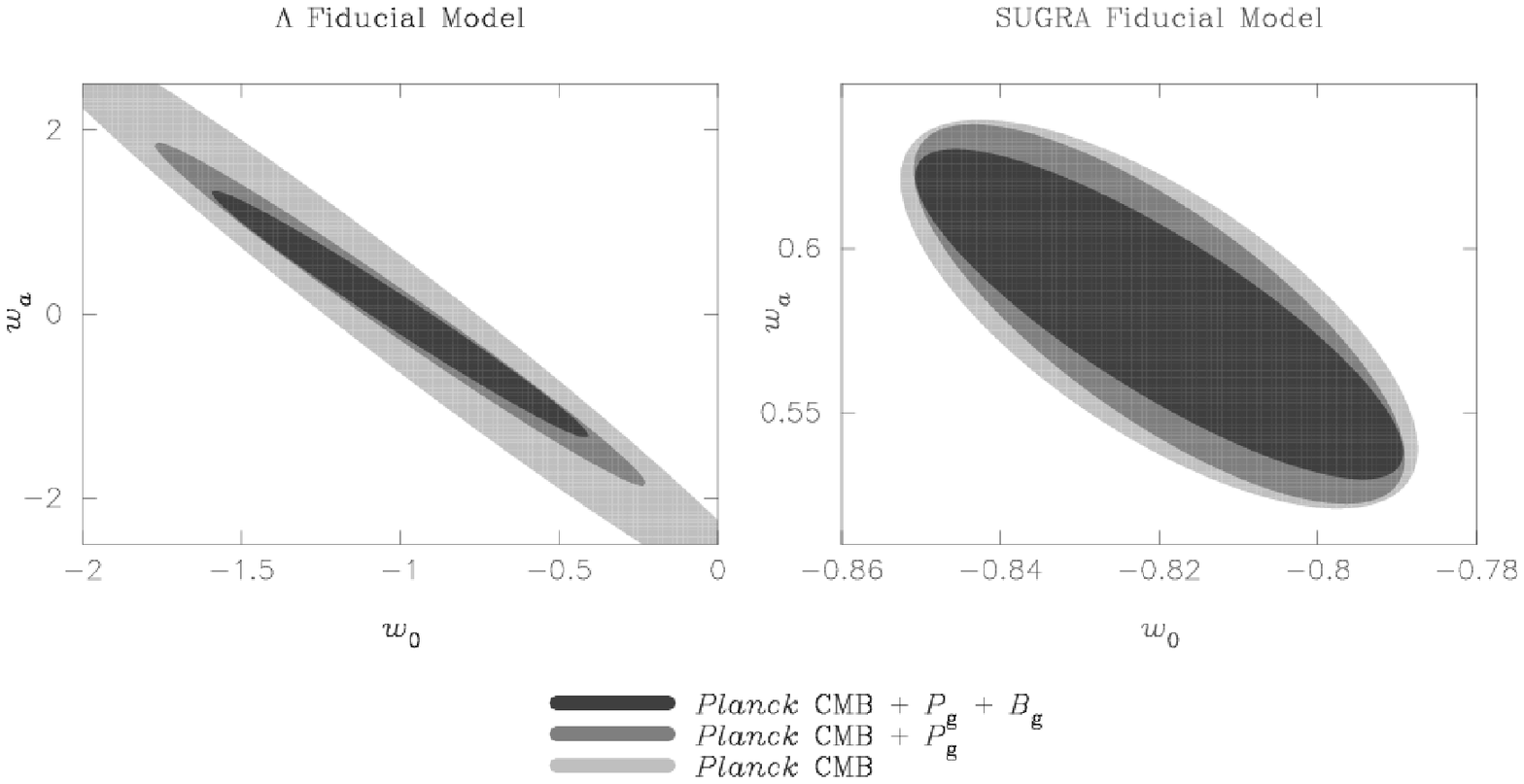}
  \caption{1-$\sigma$ constraints from the Fisher matrix analysis of
    the 11-parameter cosmological model. The $\Lambda$-model is
    compared to the SUGRA model in the two panels. Note the difference
    in the range plotted for both parameters---the SUGRA model is
    constrained much better, though the galaxy statistics add less
    information to the CMB for this model.}
  \label{fig:w_0-w_a}
\end{figure}

We performed our analysis for a supergravity (SUGRA) inspired
fiducial model, with $w_0 = -0.82$ and $w_a = 0.58$. A
non-cosmological constant fiducial model results in stronger
dark-energy constraints as a result of the redshift coverage of
galaxy observations at relatively low redshift, and CMB observations
at high redshift (\citealt{Linder2003}).
The comparison in Fig.~\ref{fig:w_0-w_a} shows an improvement
of over a factor of two in constraints on dark-energy parameters, 
though given \satellite{Planck} level CMB information galaxy clustering does not
add much. Table 2 shows these results, as well as the use of Type Ia
supernovae constraints at the level of SNAP. 
\section{Discussion} \label{sec:discussion}
We have explored the prospects for obtaining interesting dark-energy 
constraints from the angular galaxy power spectrum and bispectrum. 
The use of the bispectrum adds some new information, though the
signal-to-noise is significantly smaller than that of the power
spectrum. However, its value lies in making dark-energy constraints
robust against biasing features that cannot be measured with the power
spectrum alone. We have shown that using the configuration dependence
of the bispectrum, even a nonlinear, redshift-dependent bias
characterized by 20 parameters in our study, can be measured 
and marginalized over. 

The dark-energy constraints we obtain are weaker than those 
expected from Type Ia Supernovae or weak lensing. An ambitious
spectroscopic survey could also do better by using three dimensional
information (e.g. Seo \& Eisenstein 2003; Linder 2003). However it 
is worth emphasising that our
requirements of a multi-color imaging survey place no additional
burdens over what is needed for weak lensing. Surveys such as the CFH
Legacy survey are already in progress, and several larger surveys are
planned for the future. We have made conservative choices of redshift
binning and the maximum redshift used, to ensure that photometric
redshift uncertainties would be well below the requirements. 
Thus we believe that, at the very least, the angular galaxy
statistics would provide a useful independent check on dark-energy
constraints from imaging surveys. As discussed below, it is possible
to improve the constraints to levels competitive with the methods
mentioned above by increasing survey area, depth and quality of 
photometric redshifts. 

High redshift ($z \sim 1$) galaxies are
important to obtain the parameter constraints at the level advertised
here. This is true for several reasons. The linear regime extends to 
smaller scales at higher redshifts because nonlinear 
evolution progresses to larger scales over time. This is
evident from Fig~\ref{fig:peaks}, which demonstrates that, at
higher redshifts, a greater number of the baryon-induced peaks lie 
in the linear region. The
three dimensional nonlinear scale, $k_\rmn{max}$, is a factor of two 
smaller at $z=1.3$ than at
$z=0.3$. Further, the two-dimensional nonlinear scale,
$l_\rmn{max}$, depends not only on $k_\rmn{max}$, but also on the
angular diameter distance, which for $z=1.3$ is a factor of 3 larger
than at $z=0.3$. For ongoing and future deep imaging surveys
the number density of galaxies in these higher redshifts bins is
larger. This is helpful in that we can select special classes of
galaxies with useful properties, such as the luminous red galaxies
studied from the Sloan Digital Sky Survey. 
Even if they are a small fraction of the galaxy population, 
the shot noise contribution is not likely to be a limiting factor
as discussed above and by Seo \& Eisenstein (2003). 

\begin{figure}
  \centering
  \includegraphics{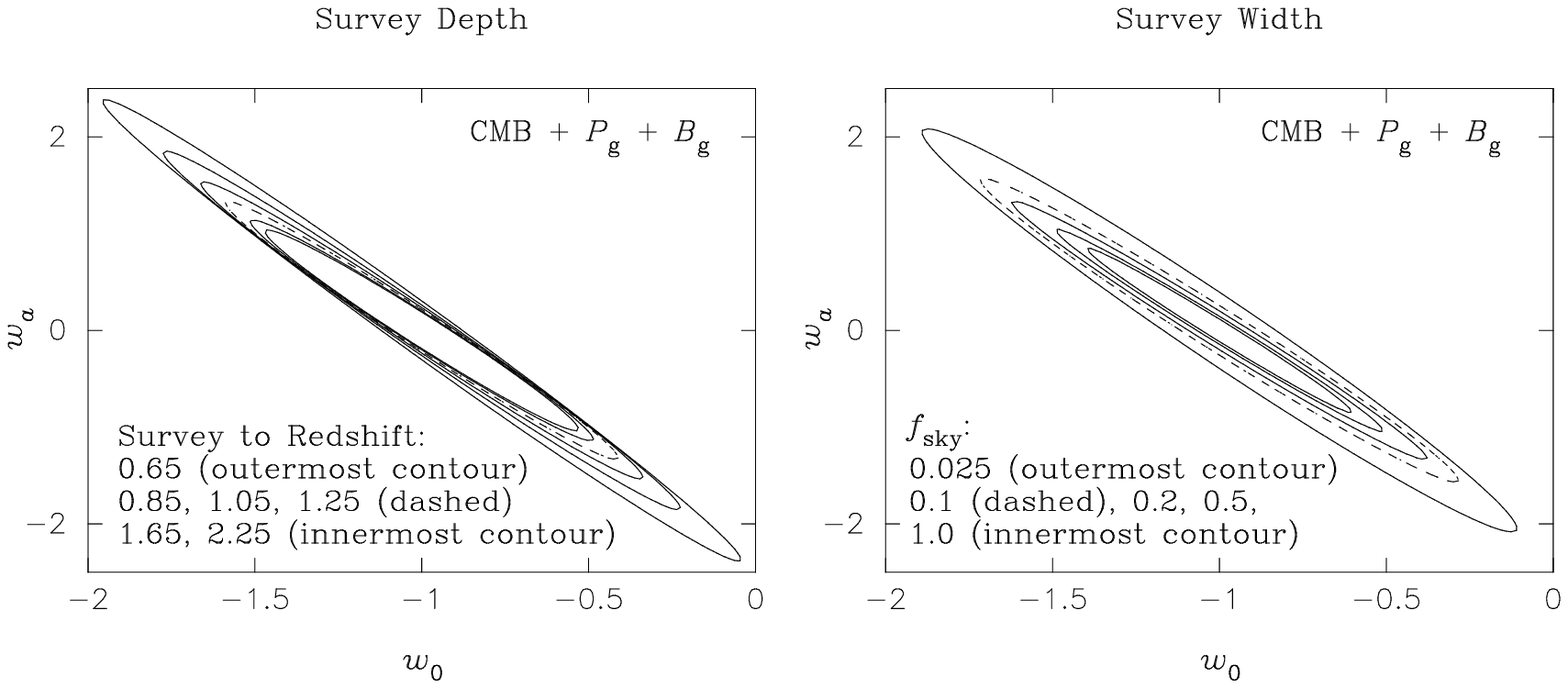}
  \caption{The two panels show the dependence of dark-energy
    constraints on survey depth and area. The left panel shows 
    constraints in the $w_a$-$w_0$ plane for different values of the 
    maximum redshift bin used for the analysis. The right
    panel shows constraints for different survey areas, given by the
    values of the sky fraction, $f_\rmn{sky}$.}
  \label{fig:z_f_sky}
\end{figure}

\begin{figure}
  \centering
  \ifthenelse{\boolean{grey-scale}}
             {\includegraphics{redshift_scaling_grey}}
             {\includegraphics{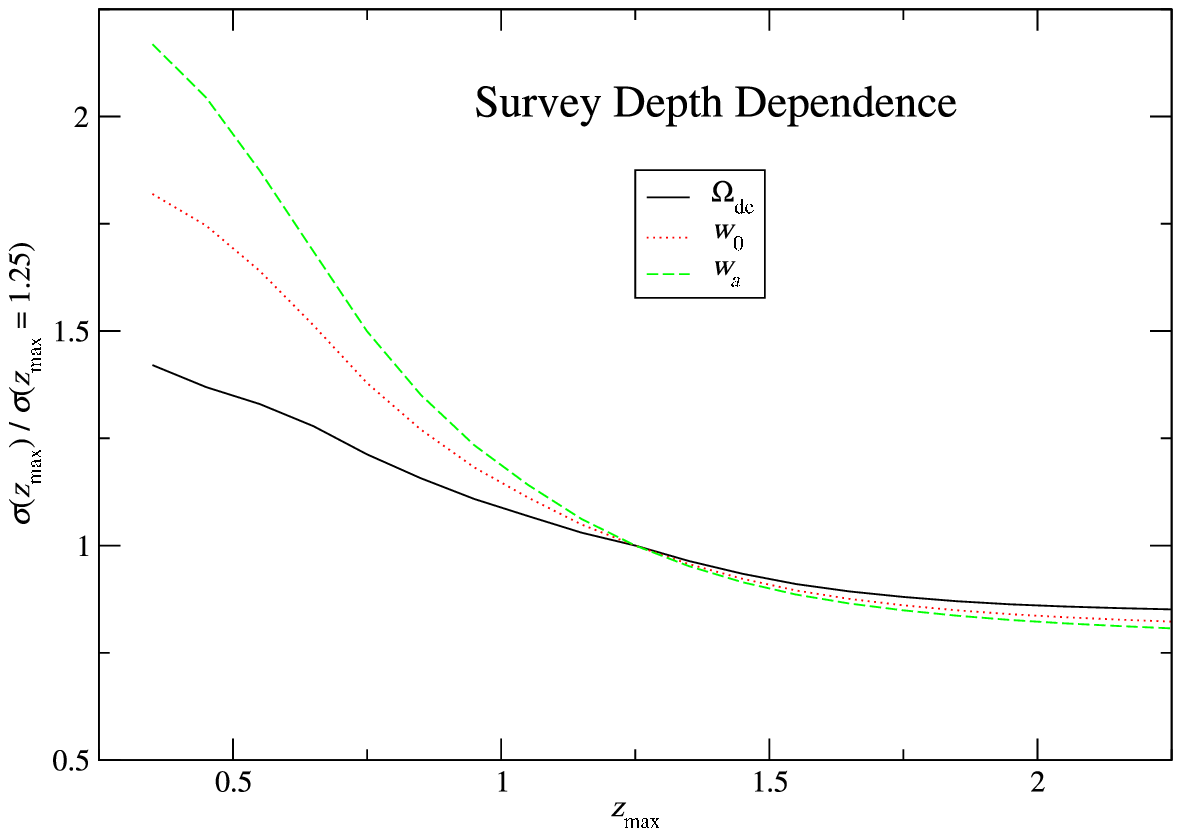}}
  \caption{Dark-energy parameter constraints as a function of the depth of the
    galaxy survey. These use the galaxy power spectrum and
    bispectrum combined with CMB data at the level of \satellite{Planck}.
    The curves are normalized to the constraints expected for a survey
    with nominal depth of $z = 1.25$.}
  \label{fig:redshift_scaling}
\end{figure}

Since we wish to learn about the time dependence of the dark-energy 
equation of state, it is important to sample our statistics over a 
wide enough range in redshift. In particular, for 
the equation of state we have used, which is linear in the scale
factor, the degeneracy between $w_0$ and $w_a$ is broken by sampling 
galaxies over several redshift bins across a range of redshifts. This
is illustrated in Fig.~\ref{fig:z_f_sky} and 
Fig.~\ref{fig:redshift_scaling}, where the
marginalized 1-$\sigma$ constraints for the dark-energy parameters are
plotted as a function of survey redshift depth. 
The constraints from a survey that extends to $z\simeq 2$ are 
twice as good as one that goes to $z\simeq 0.5$. 

The right panel of Fig.~\ref{fig:z_f_sky} shows the 
dependence of dark-energy constraints on survey area. 
While the errors do not scale simply with 
$f_\rmn{sky}$ due to the use of CMB priors, there is a significant
improvement with every factor of two in sky coverage. Using
Fig.~\ref{fig:z_f_sky}, our fiducial values can be scaled to a 
range of survey depths and areas. 

\begin{figure}
  \centering
  \ifthenelse{\boolean{grey-scale}}
             {\includegraphics{z_mean_scaling_grey}}
             {\includegraphics{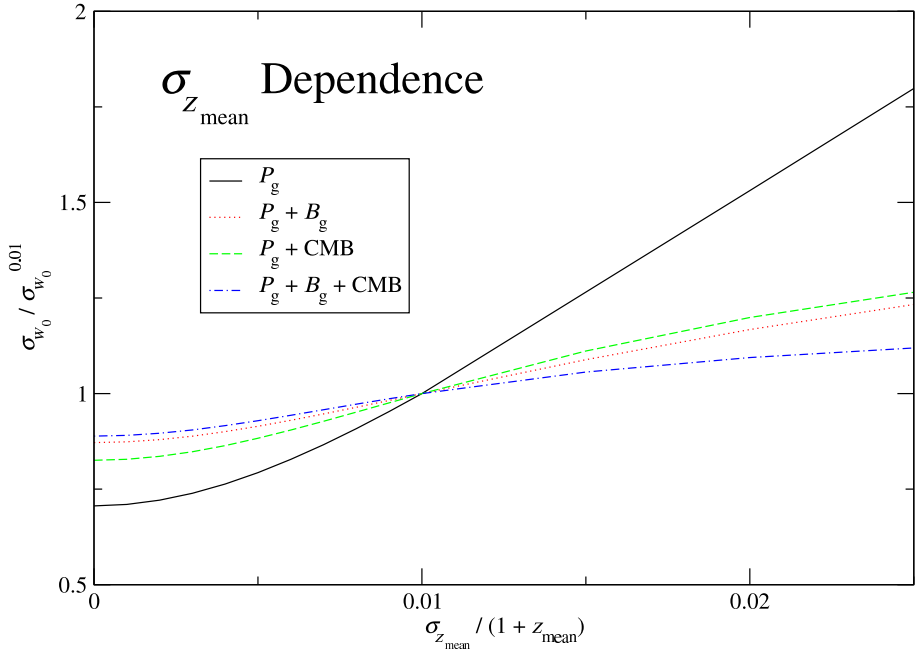}}
  \caption{Dependence of the $w_0$ constraint on the uncertainty in the
    redshift bin centers. Plotted is $\sigma_{w_0}$ normalized to the fiducial
    value (rms error of $0.01$ in $1+z$). 
    These use the galaxy power spectrum and
    bispectrum combined with CMB data at the level of \satellite{Planck}.}
  \label{fig:z_mean_scaling}
\end{figure}

The ability to use the deep imaging capability of future surveys
to go well beyond $z\sim 1$ will depend in part on the accuracy of
the photometric redshifts. With low enough scatter in these
and even a very small sub-sample of spectroscopic redshifts for 
calibration, the methods we have explored can provide much better
constraints on dark-energy models, especially models with a larger
component of the dark-energy density at higher redshifts. As discussed
by Seo \& Eisenstein (2003), the requirements on photometric redshift
scatter are not severe: an rms of $0.04$ in $1+z$ is adequate. Here
we have further checked that residual biases in the mean redshift of
each bin do not significantly degrade dark-energy constraints (see
Fig.~\ref{fig:z_mean_scaling}). Hence it is not unreasonable to hope
that the innermost contour of Fig.~\ref{fig:z_f_sky}, corresponding to
the highest redshift bin at $z=2.25$, will be achievable for a high
quality survey. In addition, a smaller rms in the photometric redshifts
will mean that we can use effectively smaller redshift bins to 
improve the constraints. A rigorous analysis of redshift binning would
require the use of cross-spectra and their covariances; alternatively, 
the galaxy distribution could be treated as three dimensional, with 
large errors in the redshift coordinate (Seo \& Eisenstein 2003). 

Besides the baryon oscillation features, dark-energy 
parameters affect the overall shape and amplitude of the angular 
power spectrum and bispectrum. To separate the effects of the 
oscillations, we calculated galaxy Fisher
matrices using the ``no-wiggles'' transfer function of
\cite{EisensteinHu1998}. If we use just the galaxy clustering
information, the constraints for the no-wiggles case are 
about twice as large. With CMB priors, the effect is smaller 
as the CMB enables the amplitude of the spectra to also provide
information. We do still have a standard ruler
in the no-wiggles case: the peak in the power spectrum
corresponding to matter-radiation equality (\citealt{CoorayEtAl2001}).
The physical length scale
of this ruler is set at decoupling by $\omega_m $, which
can be well constrained by the CMB. Hence the broad shape of the power spectrum
adds information on angular diameter distances. The amplitude depends
on the bias parameter and the growth factor, which is sensitive to
the dark energy. Using both the galaxy power spectrum and 
bispectrum with the CMB, the scalar amplitude $A_\rmn{s}$, the
two bias parameters, $b_1, b_2$, and the power spectrum 
amplitude in each redshift bin, are constrained sufficiently well. 

We have tested our choices of fiducial bias parameters and
the quasi-linear regime. We repeated our analysis by setting 
$l_\rmn{max}$ to half the value used, and find that it leads
to a 6--11 per cent degradation in the dark-energy constraints for
galaxy plus CMB information.
Thus our results are not very sensitive to the choice of cutoff of
the quasi-linear regime and the neglect of non-Gaussian contributions 
to the covariances. 

Since one of the goals of our formalism is to allow for a 
nontrivial biasing of the galaxies, we also performed our 
analysis with $b_2=1$ as the fiducial value, thus allowing
for a nonlinear term in the biasing. 
It led to improved constraints on most parameters due to a larger
impact of adding bispectrum information, e.g. the 1-$\sigma$
uncertainty in the dark-energy parameters was lower by 10 per cent. 
More importantly, we find that $b_2$ can be measured sufficiently 
well for a wide range of fiducial values, so that an effectively 
nonlinear, scale-dependent bias can be 
consistently incorporated in our analysis. This validates our 
conclusion that bias and dark-energy constraints can be simultaneously
obtained by using the bispectrum. An alternate approach to galaxy
biasing and other properties is to use the halo model, which may
also enable us to probe smaller scales, as discussed in
\cite{HuJain2003}. The analytical models for the
bispectrum need to checked and calibrated with $N$-body simulations. 
A large number of independent realizations would be needed 
since the covariances in the bispectra need to be accurately
measured as well. 
\section*{Acknowledgements}
We thank Gary Bernstein, Daniel Eisenstein, Wayne Hu, Martin
White, and Alex Szalay for helpful discussions. We are grateful to
Eric Linder for discussions and for providing his
supernovae Fisher matricies. This work is supported in part by 
NASA grant NAG5-10924 and NSF grant AST03-07297.
\label{lastpage}
\end{document}